\documentstyle[11pt,epsfig]{article}
\newcommand{\blankline}{\vskip .3cm}
\newcommand{\f}{\begin{equation}}
\newcommand{\ff}{\end{equation}}
\begin{document}
\centerline{\LARGE Causal evolution of spin networks}
\blankline
\rm
\centerline{Fotini Markopoulou${}^{\dagger *}$ and Lee Smolin${}^*$}
\blankline
\centerline{\it ${}^\dagger$ Theoretical Physics Group, 
Blackett Laboratory}
\centerline{\it Imperial College of Science, Technology and Medicine}
\centerline{\it London SW7 2BZ}
\blankline
\centerline{\it ${}^*$ Center for Gravitational Physics and Geometry}
\centerline{\it Department of Physics}
 \centerline {\it The Pennsylvania State University}
\centerline{\it University Park, PA, USA 16802}
 \vfill
\centerline{February 2, 1997}
\vfill
\centerline{ABSTRACT}
 A new approach to quantum gravity is described which joins
the loop representation formulation of the canonical theory 
to the causal set formulation of the path integral.  
The theory assigns
quantum amplitudes to special classes of causal sets, which
consist of spin networks representing quantum states of the 
gravitational field joined together by labeled null edges.  The theory
exists in $3+1$, $2+1$ and $1+1$ dimensional versions, and may
also be interepreted as a theory of labeled timelike surfaces.  The  
dynamics is specified by a choice of functions of the labelings
of $d+1$ dimensional simplices,which represent elementary
future light cones of events in these discrete spacetimes.
The quantum dynamics thus respects the discrete causal structure of 
the causal sets.    In the $1+1$ dimensional case the theory is closely
related to directed percolation models.  In this case, at least,
the theory may have critical behavior associated with
percolation, leading to the existence of a classical limit.

\blankline
email addresses: f.markopoulou@ic.ac.uk,   smolin@phys.psu.edu
\eject

\section{Introduction}

One of the oldest questions in quantum gravity is how the
causal structure of spacetime is to be preserved in a quantum
theory of gravity in which the metric and connection fields
are expressed as quantum operators.  As argued by Roger
Penrose some time ago\cite{roger-worry}, if 
the metric of spacetime is subject
to quantum fluctuations then the causal structure will become
uncertain, so that there may be some nonvanishing amplitude
for information to propagate between any two spacetime events.
But in this case it is not clear what the canonical
commutation relations could mean as they are defined with
respect to an {\it a priori} causal structure.  Clearly this is the sort
of problem that can only be resolved within the context
of a complete and physically sensible quantum theory of
spacetime geometry.

The solution proposed by Penrose is that the causal structure
should stay sharp while the notion of spacetime points or events
become indistinct\cite{roger-worry}.  Here we would like to propose 
a related, but
different solution to this puzzle in the context of a discrete
formulation of quantum gravity.  In this framework there are 
discrete
quantum analogues of both null rays and spacetime events.
The latter are sharply defined because they are indeed defined
in terms of the coincidence of causal processes.  Quantum amplitudes
are then defined in terms of sums over histories of discrete
causal structures\cite{causalset,thooft-cs}, each of which are 
constructed by a set of rules
that respect its own causal relations.

In this paper we realize this proposal in a class of theories of
the quantum gravitational field that 
combines the kinematical structures 
discovered through the program of canonical quantization with
a discrete causal structure that captures the main
features of the causal structure of Minkowskian spacetimes.   
To describe it we may begin by recalling the main result of
non-perturbative quantum 
gravity
\cite{abhay1,tedlee,lp1,lp2,ls-review,abhay-book,carlo-review}, 
which is the identification of
the basic states and operators of the theory.   The kinematical
state space  consists of diffeomorphism classes of 
spin networks\cite{sn1,volume1}.
These are endowed
with a geometrical interpretation by the fact that the 
spin-network basis makes possible the diagonalization of the
operators that correspond to three dimensional geometrical
quantities, such as area\cite{ls-review,volume1,AL1}, volume
\cite{ls-review,volume1,renata-volume,dpr,tt-volume,L1,rsl,RB} 
and length\cite{tt-length}.  The spectra of all of these observables 
are
discrete, which gives rise to a picture in which 
quantum geometry is discrete and combinatorial.

The spin network states and the associated operators may be 
considered a complete solution to the problem of the kinematics
of quantum general relativity at the level of spatial diffeomorphism
invariant states.  It may also be considered to have been derived
from classical general relativity through a standard and well
understood quantization 
procedure.  What is required to complete the theory
is then to specify the dynamics by which the quantum geometries
described by the spin networks evolve to give rise to quantum
spacetimes.  This is the goal of this paper.  
What we do below is to describe a set of rules that allow us to
construct four dimensional description of the evolution in time
of spin networks that is both completely non-perturbative and
realizes a precise discrete causal structure. 

As we describe below, the amplitude for a given initial spin network
state to evolve to a final one is given in terms of a sum over
a special class of four dimensional combinatorial structures,
which are called {\it spacetime networks}.    Each such structure,
which we take as the discrete analogue of a spacetime,
is foliated by a set of discrete
spatial slices, each of which is a combinatorial 
spin-network.   These discrete ``spatial slices" are then connected
by ``null" edges, which are discrete analogues of null geodesics.
The rules for the amplitudes are set up so that information
about the structure of the spin networks, and hence the
quantum state, propagates according to the causal
structure given by the null edges.  

The dynamics is specified by a set of simple rules that both 
construct the spacetime networks, given initial spin networks,
and assign to each one a probability amplitude.
Each spacetime net is then
something like a discrete spacetime. More precisely, each is a causal
set\cite{causalset,thooft-cs}.  This is a set of points which has the 
causal 
properties that may be
assigned to sets of points in a Minkowskian spacetime: to each pair
either one is to the future of the other, or they are causally 
unrelated.
Thus, our proposal may be said to resolve the problem of specifying
the dyanmics of non-perturbative states of quantum gravity
in a way that utilizes elements of the causal set picture of discrete
spacetime.

It must be emphasized that the form of dynamics 
we propose here is not
derived through any procedure from the classical theory.  
Instead, we seek the simplest algorithm for a transition amplitude
between spin network states that is consistent with some
discrete microscopic form of causality.
 The reason for this is that attempts to follow the procedure
of canonical quantization, although having led to partial 
success\cite{ham1,roumen-ham,RB,qsdi,qsdii},
face both conceptual and technical problems that it is not clear
can be resolved successfully.    Besides the problem of causal
structure mentioned above, there is the whole problem of time
and observables in quantum cosmology.  In addition, while it
seems to have been possible to construct well defined finite
diffeomorphism invariant operators that represent the hamiltonian
and hamiltonian constraint\cite{ham1,roumen-ham,RB,qsdi,qsdii}, 
these suffer from problems related
to both the algebra of quantum constraints and the existence
of a good continuum limit\cite{instability}.   

However, it may not be necessary that these problems be resolved.  
>From the
path integral point of view, the Planck scale dynamics need only
have one property to lead to a successful quantum theory of gravity,
which is that the discrete theory it gives rise to has critical behavior,
so that a good continuum limit exists in which the universe becomes
large and curvatures (suitably averaged) are 
small\cite{weinberg,fixedpoint,ll-fractal,critical}.   When this is
the case, standard renormalization group arguments guarantee that
the macroscopic dynamics will be governed by an effective action
whose leading term is the Einstein-Hilbert action\cite{weinberg}.  
Thus, the necesssary
criteria that the microscopic dynamics must satisfy is only that it 
give
rise to such critical behavior.  It is neither necessary, nor may it be
possible\cite{instability}, that a form of microscopic dynamics 
that satisfies this
condition come from a ``quantization" of general relativity.

The framework we describe here in fact gives rise to a class of
theories, which are distinguished by the amplitudes given to certain
combinatorial structures.   A key question is then whether any
of the theories in this class give rise to critical behavior.  As we will
describe below, two considerations suggest this may be possible.  
First,
the form of the path integral is close to that which arises in three and
four dimensional topological field theories\cite{4dtqft}.  
This suggests we are on
the right track, as  combinatorially
defied topological field theories have a trivial form of critical
behavior in that they have no local degrees of freedom.  It is
reasonable to conjecture 
that theories with massless degrees of freedom may be
found on renormalization group trajectories that
approach fixed points associated with topological quantum
field theories.

Second, the form of the path integral we propose is very similar to a
class of systems that has been well studied in statistical
physics,  which is directed percolation\cite{DP,MZ}.    
As we will argue, it is likely
that at least some of the theories we describe are in the universality
class of directed percolation, which means that they will have
critical behavior necessary for the existence of the classical limit.
 Then, given the fact that each network is also a causal set, it may
be possible to identify the networks which dominate in the
continuum limit with a classical spacetime, using the ideas 
previously explored for general causal sets by Bombelli et 
al\cite{causalset}
and 't Hooft\cite{thooft-cs}. 

The form of the path integral we propose is also similar to 
a recent proposal of Reisenberger and Rovelli\cite{RR}, which 
however
gives a discrete form of the Euclidean path integral.  In fact, the
direct impetus for our work was the desire for a path
integral that incorporates a discrete form of causal structure, suitable
for describing the real, Minkowskian theory, while preserving many
of the attractive features of the Reisenberger-Rovelli formulation,
such as its relationship to topological quantum field theory.  
 
The $2+1$, $3+1$ and $1+1$ theories are described, respectively,
in sections 3,4 and 5, after which the paper concludes with some
final comments and directions for future work.

\section{Kinematics of spin networks}

For the purposes of this paper a spin network is a combinatorial
labeled graph whose nodes and edges are labeled according to the
rules satisfied by spin networks\cite{sn1,sn-roger}.
The edges are labeled by representations of $SU(2)$ and the nodes
are labeled by intertwiners, which are distinct ways of extracting
the identity representation from
the products of the representations on the incident edges.  For each
node $n$ of valence higher than three there is a finite dimensional
space ${\cal V}_n$ of intertwiners, which may be labeled by
virtual networks, which are spin networks which represent the
state of the node\cite{sn1}.  These are well defined up to the recoupling
relations\footnote{For reviews of spin networks see\cite{KL,review}.}.

We will not be concerned here with additional information 
corresponding to diffeomorphism classes such as the continuous
parameters that specify higher valence nodes.  In fact, we consider
the spin networks to be defined only by their combinatorics, no
embedding in a spatial manifold is assumed.

\section{Rules for causal evolution: $2+1$ case}

As the spin networks we employ are combinatorial structures,
the dimension of space must be determined from combinatorial
information in the networks alone.  We describe two versions of
our theory, which are appropriate for $2+1$ and $3+1$
dimensional spacetime, respectively.  We begin with the
$2+1$ dimensional theory as it is easier to visualize, the 
$3+1$ dimensional version will be obtained from it by 
increasing the valences of the nodes in a particular way we
will describe.

The algorithm for causal evolution we propose consists of two rules
which are applied alternatively. 

\subsection{Rule 1}

Consider an initial spin network $\Gamma_0$, which consists of
a set of edges $e_{ij}$ and nodes $n_i$ (where $e_{ij}$
connects the two nodes $n_i$ and $n_j$).  
To obtain the $2+1$ dimensional version of the theory
we will restrict $\Gamma_0$ to be trivalent, which means it
can be embedded in a two dimensional surface.  

The first evolution
rule constructs a successor network $\Gamma_1$ together with
a set of ``null" edges which each join a node of $\Gamma_0$ to
$\Gamma_1$.  The rule is motivated by the idea that the
null edges should correspond to a discrete analogue of null
geodesics joining spacetime events.  

\begin{figure}
\centerline{\mbox{\epsfig{file=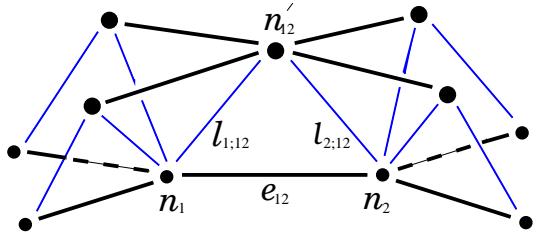}}}
\caption{The construction of a new spin network under Rule 1.}
\label{basic}
\end{figure}

To each edge $e_{ij}$ of $\Gamma_0$ we associate a node
$n_{ij}^\prime$ of the successor network $\Gamma_1$.  
We connect the new node $n_{ij}^\prime$ to  
$n_i$ and $n_j$, the nodes at the ends of $e_{ij}$ by two null
edges.  (Why they are called null will be clear below, for the
moment a null edge is just an edge connecting a node
of an initial
spin network to one of a successor under the evolution rules.) 
The null edge connecting $n_i$ of $\Gamma_0$ to
$n_{ij}^\prime$ of $\Gamma_1$ will be called $l_{i;ij}$.
(See Figure \ref{basic}).

\begin{figure}
\centerline{\mbox{\epsfig{file=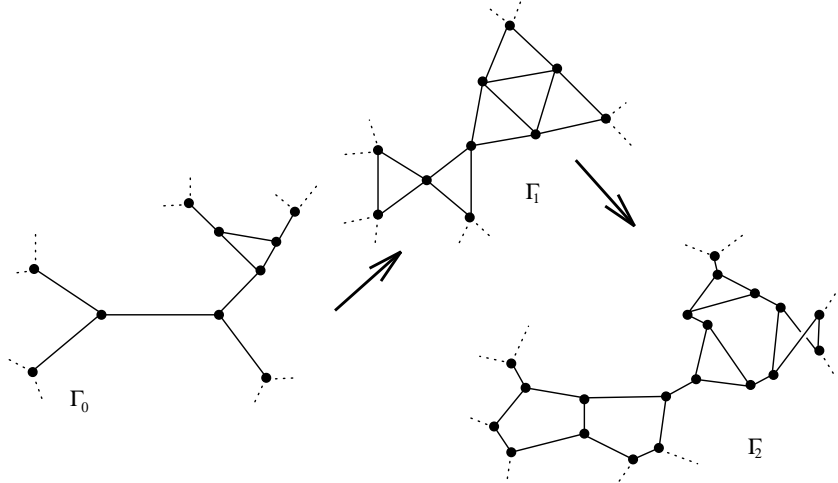}}}
\caption{Two steps in the evolution of a trivalent spin network,
following first Rule 1, then Rule 2.}
\label{graphs}
\end{figure}

Two of the
nodes of $\Gamma_1$, 
$n_i^\prime$ and $n_j^\prime$ will be connected
by an edge $e^\prime_{ij}$ in $\Gamma_1$ 
if the edges $e_i$ and $e_j$ were
incident on
a common node $n_\alpha$  in $\Gamma_0$.  The result is that
the new graph $\Gamma_1$ is related to the old one by a kind of
duality in which edges go to nodes and nodes go to complete 
graphs, which are graphs in which each node is connected
to every other node. (See Figure \ref{graphs}).

The result of this rule is a spacetime spinnetwork ${\cal G}_{01}$
bounded by the two ordinary spin networks $\Gamma_0$ and
$\Gamma_1$ whose nodes are connected by a set of null edges.
In general a spacetime spin network (or spacetime net, 
for short) will consist of a set of $N$
ordinary spin network, $\Gamma_i$, $i=0,1,...,N$, together
with a set of null edges that join nodes of $\Gamma_i$ to
nodes of $\Gamma_{i+1}$.  We may also have need to refer to
the graph made only of the null edges that join the spinnets,
which we call the internet.

\begin{figure}
\centerline{\mbox{\epsfig{file=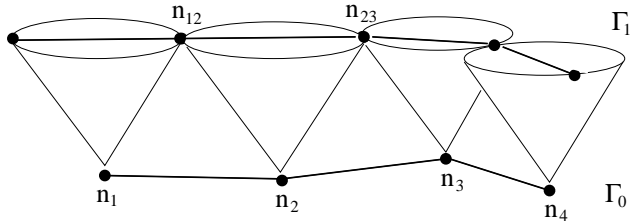}}}
\caption{The new nodes represent events defined by meeting of causal 
processes.}
\label{cones}
\end{figure}

The motivation for this construction is the following:  We imagine
that the initial spin network $\Gamma_0$ 
is embedded in a spacelike slice
$\Sigma_0$ of a four dimensional spacetime ${\cal M}$, such that
the edges correspond to spacelike geodesics.  (This imaginary
embedding is only for motivation, once the rules are 
established it plays no role.)
Each node $n_i$ of $\Gamma_0$ emits a light signal which
evolves into its future lightcone in $\cal M$.  For each 
$n_i$ and $n_j$ connected by an edge $e_{ij}$, 
there will be an event at which the light
rays from those nodes, each traveling in the direction of the
geodesic to the other edge, first meet.  This event corresponds
to the new node $n^\prime_{ij}$ and the null edges
$l_{i;ij}$ and $l_{j;ij}$ correspond to the null rays emmited
by $n_i$ and $n_j$ that met at $n_{ij}^\prime$.
(See Figure \ref{cones}.)

For each edge $e_{ij}$ of $\Gamma_0$ we then have an event
to the future, $n_{ij}^\prime$.  These are the nodes of the
successor spin network $\Gamma_1$; we may imagine that they
are embedded in a second two dimensional surface
$\Sigma_1$ embedded in $\cal M$ to the future of $\Sigma_0$.
(See Figure \ref{cones}).  We will assume that this surface can be
chosen so that it is spacelike, this corresponds to the fact that
in the construction there are no null edges that connect
nodes in the same $\Gamma_i$. (Of course whether this can
be done in an arbitrary spacetime $\cal M$ given an
arbitrary spacetime metric $g_{ab}$  is a dynamical question,
but as we are specifying the dynamics  and the spacetime
metric in terms of the discrete
structure there is no loss of generality in assuming this.  To
put this another way, as 
the embedding is only for motivation, we need only that it is possible
to choose some metric $g_{ab}$ such that the surface
$\Sigma_1$ is spacelike.)

The basic idea of the construction is that 
the rule for assigning edges that connect the new nodes 
$n_{ij}^\prime$, as well
as the rules for assigning amplitudes to labelings of the edges
must satisfy a discrete causal principle.  This causal
principle is stated in terms of a discrete  causal 
structure, which is  specified by the  
null edges.  Thus, each node in the spacetime net 
$\cal G$ has a future and past light cone, which is gotten
by following null edges from it to the past or the future.  Two
nodes of a spacetime net $\cal G$ will be
said  to be causally connected if and only
if there is a path
of future pointing null edges,  that link one to the other.  
The causal past or future of
a node then consists of all nodes to which it is causally connected
by a path of null edges going into the past or future, and all
edges such that both ends are in the causal past or future of it.  
The causal
past or future of an edge is the union of the 
causal past or future of its two ends.
Given this structure we propose
a {\it principle of discrete causality} which 
says that: 

\begin{itemize}

\item{}The information about
which other nodes a node is connected to, as well as the colorings
of a node or edge, can only be determined by information
in its causal past, except that the assignment of amplitudes
may induce correlations among two edges that share a common
node.

\end{itemize}

In this definition, the information available at a node
is its color and the number and colors of the edges incident on it;
the information available at an edge is its color and the colors
of its ends. 

\begin{figure}
\centerline{\mbox{\epsfig{file=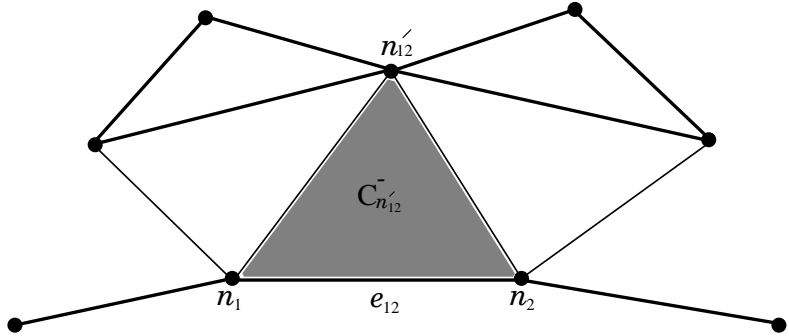}}}
\caption{The backwards light cone ${\cal C}_{n^\prime_{12}}^-$ of the node
$n^\prime_{12}$.}
\label{lightcone}
\end{figure}

Thus, which other nodes $n^\prime_{kl}$ a given
node $n^\prime_{ij}$ may be connected to in 
$\Gamma_1$ can be determined only by information that is in
the backwards light cone of $n^\prime_{ij}$,
which will be denoted ${\cal C}_{n^\prime_{ij}}^-$
(See Figure \ref{lightcone}.)
In addition to this we will make a second assumption, which
is that the dynamics be as local in time as possible.  This means
that the information necessary to specify the connectivity 
of a node in a successor spinnet $\Gamma_1$, or the color
of one of its edges should depend directly
only on information available in the
intersection of the backwards light cone of that node
or edge with the previous spinet $\Gamma_0$,
and not on any information from earlier spin networks.  

The only part of $\Gamma_0$ that is in 
${\cal C}_{n^\prime_{ij}}^-$ consists of the two nodes  
$n_i$ and $n_j$ and the edge $e_{ij}$ that joins them.  
(See Figure \ref{lightcone}).
Therefore
all the information that determines who $n^\prime_{ij}$ is connected
to and how  it is labeled must be available there.  This information 
consists of the labeling
of $e_{ij}$,  the information about which other edges
of $\Gamma_0$ are incident on $n_i$ and $n_j$ and the
labelings of those edges.

The simplest rule for connecting the nodes of $\Gamma_1$ 
consist with this discrete causality principle is the one we have
given: two nodes are connected if the edges they correspond
to in $\Gamma_0$ were incident on the same node.

We may note that no instruction is given for how the
new spinnet $\Gamma_1$ may be embedded in a two
dimensional manifold $\Sigma_1$.  The spinnets used here
are to be considered to be purely combinatorial structures, which
come with no such embedding information.  This is true as well
for the spacetime spinnets ${\cal G}$.  While we may use a
picture in which $\cal G$ is embedded in some $2+1$ dimensional
spacetime, this is only to allow us to use our intuition about
causal structure to motivate the rules and principle of causality
for the discrete construction.  Once the construction is specified the
notion of a spacetime continuum may be recovered only in the
case that the dynamics shows critical behavior that allows us
to define a continuum limit.

\begin{figure}
\centerline{\mbox{\epsfig{file=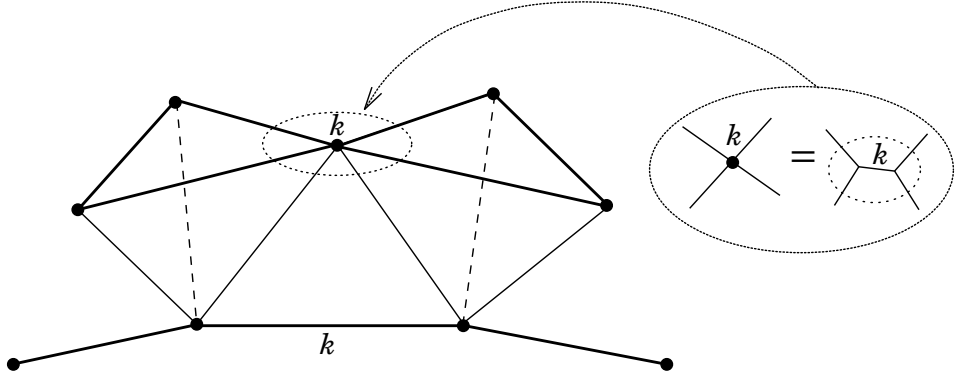}}}
\caption{The labeling of a new four valent nodes is gotten from 
that of the edge that gave rise to it. A dotted circle around 
a subgraph means that it is virtual.}
\label{k}
\end{figure}

We have yet to specify the labelings of the nodes and edges
of $\Gamma_1$.  To do this let us recall that the original
graph $\Gamma_0$ was trivalent. Then each node
 in the successor graph
$\Gamma_1$ is four valent  (See Figure \ref{graphs}). 
There is a natural choice of
assignment of its state, which is that it is given by
a virtual spin network in which the four valent node
is decomposed into two trivalent nodes joined by
a virtual edge parallel to $e_{ij}$ (See Figure \ref{k}).
The virtual edge  can then be colored  by
the same spin that labels $e_{ij}$.

There is no natural unique assigment for the labelings 
of the edges of $\Gamma_1$.  Instead we will assign
a complex amplitude ${\cal A}_{0 \rightarrow 1}$ to
each set of labelings of the edges of $\Gamma_1$,
given the labelings of $\Gamma_0$.  To see how to do this
we note that there are two restrictions on the labelings and
amplitudes.  The first is that the labelings of the edges of
$\Gamma_1$ must be consistent with the labelings on the
nodes, which have already been determined.  This condition is
easily stated in terms of the decomposition of each node of
$\Gamma_1$ into trivalent virtual nodes, each then has
one incident virtual edge whose labeling has been determined
from the previous paragraph and two real edges whose labeling
must be determined, the possible labelings of the real edges
must then be chosen so that the addition of
angular momentum is satisfied at each virtual node.

\begin{figure}
\centerline{\mbox{\epsfig{file=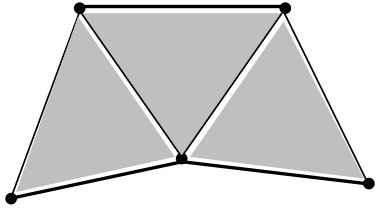}}}
\caption{The  backwards light cone of an edge.}
\label{edge}
\end{figure}

The second restriction is the principle of causality.  The discrete
backwards lightcone of an edge $e^\prime_{(ij)(ik)}$ in
$\Gamma_1$ contains a subgraph $\gamma $ of 
$\Gamma_0$ that contains
the nodes $n_i$, $n_j$ and $n_k$ as well as  the two edges
$e_{ij}$ and $e_{ik}$.  (See Figure \ref{edge}).  

\begin{figure}
\centerline{\mbox{\epsfig{file=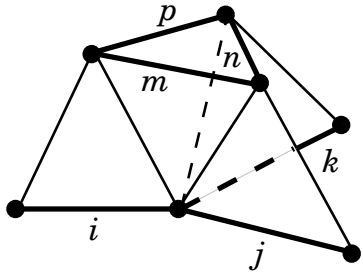}}}
\caption{The evolution of a trivalent node.}
\label{node}
\end{figure}

Therefore the information
relevant for the labeling of $e^\prime_{(ij)(ik)}$ must be
taken from only information available at those nodes and
edges.  A perscription consistent with this restriction is the
following. Consider the three pairs of edges incident
at a node $n_i$, which are $e_{ij}, e_{ik}$ and $e_{il}$.
These have labelings that for simplicity we may denote
by $i,j,k$ and $l$.  The node $n_i$ gives rise by the evolution
rule to a triangle whose edges are (See Figure \ref{node}),
$e^\prime_{(ij)(ik)}$, $e^\prime_{(ij)(il)}$  and
$e^\prime_{(ik)(il)}$. For simplicity let us denote these
edges simply by $e^\prime_m$, $e^\prime_n$
and $e^\prime_p$ and their labeling by $m,n$ and $p$,
respectively.  The simplest assumption consist with the
restriction of causality is that there is an amplitude
$J(mnp; jkl)$ for each choice of labeling of the new edges
$mnp$, given the labelings of the old edges $ijk$.

\begin{figure}
\centerline{\mbox{\epsfig{file=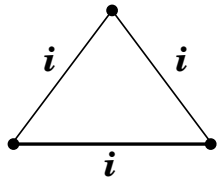}}}
\caption{The  label of a null edge is the same as that of the spacelike
edge that gave rise to it.}
\label{iii}
\end{figure}

To see what conditions this amplitude $J(mnp; jkl)$ must
satisfy, let us note that it corresponds naturally to a 
tetrahedron in the spacetime network 
${\cal G}_{01}$.  One thing we have not done
is labeled the null edges.  However, there is a natural way
to do this, which is to note that each edge $e_{ij}$ in
$\Gamma_0$ gives rise to two null edges $l_{i;ij}$
and $l_{j;ij}$ in the spacetime network.   The intersections
of their causal pasts with $\Gamma_0$ consists only of
the edge $e_{ij}$ and the two nodes it joins.  However, as the
nodes of $\Gamma_0$ are assumed to be trivalent and, hence,
unlabeled, the only information a labeling of the two null edges
could depend is the color of the original edge.  Therefore, the
only natural assumption is that the null edges are labeled with the
same coloring as the edge it came from (See Figure \ref{iii} ).

\begin{figure}
\centerline{\mbox{\epsfig{file=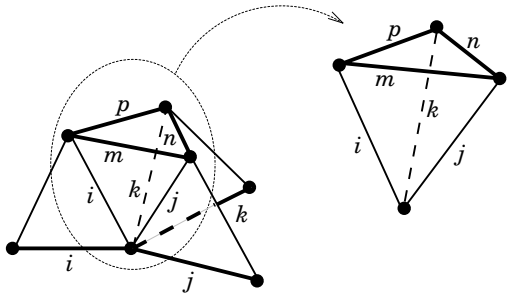}}}
\caption{The spacetime tetrahedron used to determine the labels $m,n,p$
on the edges of a triangle that is the future evolution of a
trivalent node.}
\label{tetrahedron}
\end{figure}

There is then a tetrahedron ${\cal T}_i$ in 
${\cal G}_{01}$ corresponding to each node $n_i$ of
the original spin network $\Gamma_0$ (see figure \ref{tetrahedron}).  
It contains three 
null edges, whose labelings are known and three new spacelike
edges which are part of the new spin network $\Gamma_1$.
There is then an amplitude $J(mnp;jkl)$ associated to each such
tetrahedron.  This amplitude must be consistent with a subset of the
symmetries of the tetrahedron, which are those that do not mix
the null and spacelike edges.

The total amplitude will then be taken to be
\f
{\cal A}_{0 \rightarrow 1} = \prod_i J(mnp;jkl)
\ff
where the product is over all nodes $n_i \in \Gamma_0$.

The choice of a function $J(mnp;jkl)$ corresponds to the choice
of dynamics.  The parameter space of the theory then
consists of the possible functions $J(mnp;jkl)$ of
the six spins in Figure 9 that is invariant under rotations in
space of the spacelike triangle.  Within this space
of possible theories is a special one, based on the choice,
\f
J(mnp;jkl)= {\cal T}[mnp;jkl ]
\ff
where  ${\cal T}[mnp;jkl ]$ is the ``tetrahedronal symbol", 
which is a $6j$ symbol normalized so that it has all the
symmetries of the tetrahedron\cite{KL}.
The ${\cal T}[mnp;jkl ]$ 
has more symmetry than we need and is also special in that
it satisfies the $6j$ symbol identities.   It is possible that
with  this 
choice the theory corresponds
to a $2+1$ dimensional 
topological quantum field theory, but that has so far not been shown.

\subsection{Rule 2}

We might just apply Rule 1 over and over again, but the result would
be that each successor spin network has nodes of higher and higher
valency.  (This is easy to see, if each node of $\Gamma_n$ has 
valence $P$, each node of $\Gamma_{n+1}$ will have valence
$2(P-1)$.)  To prevent this from happenning we need a second
rule that  lowers rather than raises the valence of the nodes.
There is a natural choice for such a rule which is the following.
Recall that each higher-than-trivalent node $n_i^\prime$ 
has associated to it a state $| \eta \rangle$ in a finite dimensional state
space.  This space,  ${\cal V}_{n_i}$, is spanned by
a basis of states $| \gamma \rangle$, 
each of which can be represented as an open
spin network whose ends are the edges incident on $n_i$
connected to each other through a set of virtual trivelent nodes
and edges.  
In the case of a four valent node, the space may be labeled
${\cal V}_{4,ijkl}$, where $i,j,k$ and $l$ are the spins of the
four edges incident on it.

\begin{figure}
\centerline{\mbox{\epsfig{file=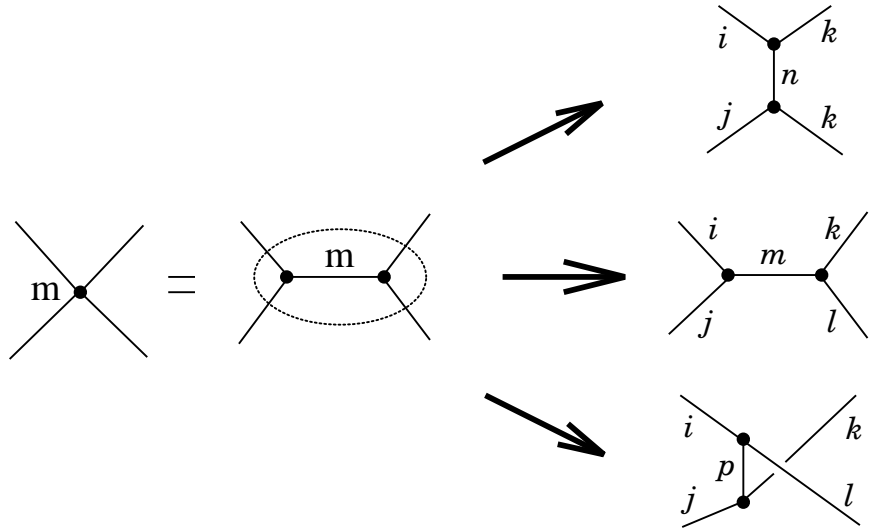}}}
\caption{The three ways a node splits under Rule 2.}
\label{msplitting}
\end{figure}

A four valent network can be labeled by inserting a virtual edge,
as we have already indicated in Figure 5.  For each node
in $\Gamma_1$ there is a natural way to split it, parallel to
the edge that gave rise to it.  Corresponding to that we 
may evolve the network by splitting it, so that the
virtual node becomes real (See the middle term in Figure \ref{msplitting}.)

\begin{figure}
\centerline{\mbox{\epsfig{file=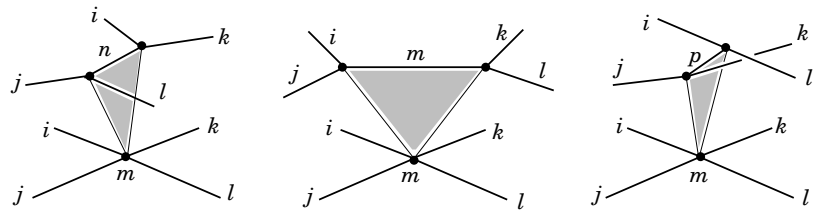}}}
\caption{The effect of the three splittings in Rule 2.}
\label{directterm}
\end{figure}

The effect of this shown in Figure \ref{directterm}.

The new edge created is labeled by the same label $m$ that
was on the edge in $\Gamma_0$ that gave rise to the node
in $\Gamma_1$.  

However, there are two other ways that the node could be split
into a pair of trivalent nodes.  Let us call the first way the
``s channel", and the other two the ``t channel" and ``u channel",
by analogy to scattering theory (See Figure \ref{msplitting}).  
Associated with each 
there are states in ${\cal V}_{4,ijkl}$, which may be labeled
$|t;n\rangle$ and $|u;p\rangle$.  (The original state shown in 
Figures 5 is then called 
$|s;m\rangle$.)  Each of these may be split, giving rise to two new
trivalent nodes and a new edge, which has the same label
as the virtual edge it came from.

Rule 2 may then be stated as follows:  

\begin{itemize}

\item{}$\Gamma_2$ 
consists of a sum of terms which are gotten from $\Gamma_1$
by splitting each four valent node in each possible way
corresponding to the $s,t$ and $u$ chanel states in the
spaces ${\cal V}_{4,ijkl}$.  The $s$ channel split is
multiplied by an amplitude $a$.  Each $t$ channel
split is multiplied by an amplitude $b\langle t;n|s;m\rangle$ and
each $u$ channel split by $c \langle u;p|s;m\rangle$ where
$|a|^2 + |b|^2 + |c|^2 =1$.

For each spin network $\Gamma_2$ produced by the rule the 
amplitude ${\cal A}_{\Gamma_1 \rightarrow \Gamma_2}$ is then
the product of these factors for every four valent node that is
split.
 
\item{}In each term 
each of the two new nodes are then connected to the original
node by a new null edge, as in Figure 10.   
The two null edges created may be
labeled by the same label as the new spacelike edge associated
with them.

\item{}Rule 2 also preserves all of the edges of $\Gamma_1$, which
appear in $\Gamma_2$, with the same labels.

We may note that there is freedom in the specification of Rule
2, associated with the choices of the amplitudes $a,b,c$ for
the $s,t$ and $u$ channels.  Unless they are needed, however,
it is simplest to set them equal so that $a=b=c=1/\sqrt{3}$.

\end{itemize}

\subsection{Combining the two rules: the transition law}

The effect of Rule 2 is to make the resulting graph $\Gamma_2$
trivalent. If we apply the two rules in succession, starting with
any initial spinnet, we generate a discrete causal graph
${\cal G}$ which is foliated by spin networks that are alternatively
trivalent and four valent (here and elsewhere in this paper,  
valence counts only spacelike edges and ignores the null edges).  
The nodes of these spinnets are connected by null edges
in the following way: each trivalent node has one null edge going
into the past and three going into the future.  Each four valent node
has two null edges going into the past and two going into the future.

We may then state the dynamics of quantum gravity in the following
form:  Given two trivalent spin networks $\Gamma_i$ and
$\Gamma_f$ we construct the amplitude 
${\cal A}_{\Gamma_i \rightarrow \Gamma_f }$ for the 
first to evolve to the second.  We consider all causal 
spacetime nets
${\cal G}$ consistently built by the alteration of the two rules
which have $\Gamma_i$ as the zeroth spin network and 
$\Gamma_f$ as the last.  $\cal G$ will have
an odd number $L_{\cal G}$ of component spinnets,
$\Gamma_I$.  
We then have
\f
{\cal A}_{\Gamma_i \rightarrow \Gamma_f }= 
\sum_{\cal G} \prod_{I=0}^{L_{\cal G}-1}  
{\cal A}_{\Gamma_{I} \rightarrow \Gamma_{I+1}}
\ff
where the sum includes the sums over all the allowed colorings
and the amplitude is defined alternatively in terms of Rule 2 or
Rule 1.

\begin{figure}
\centerline{\mbox{\epsfig{file=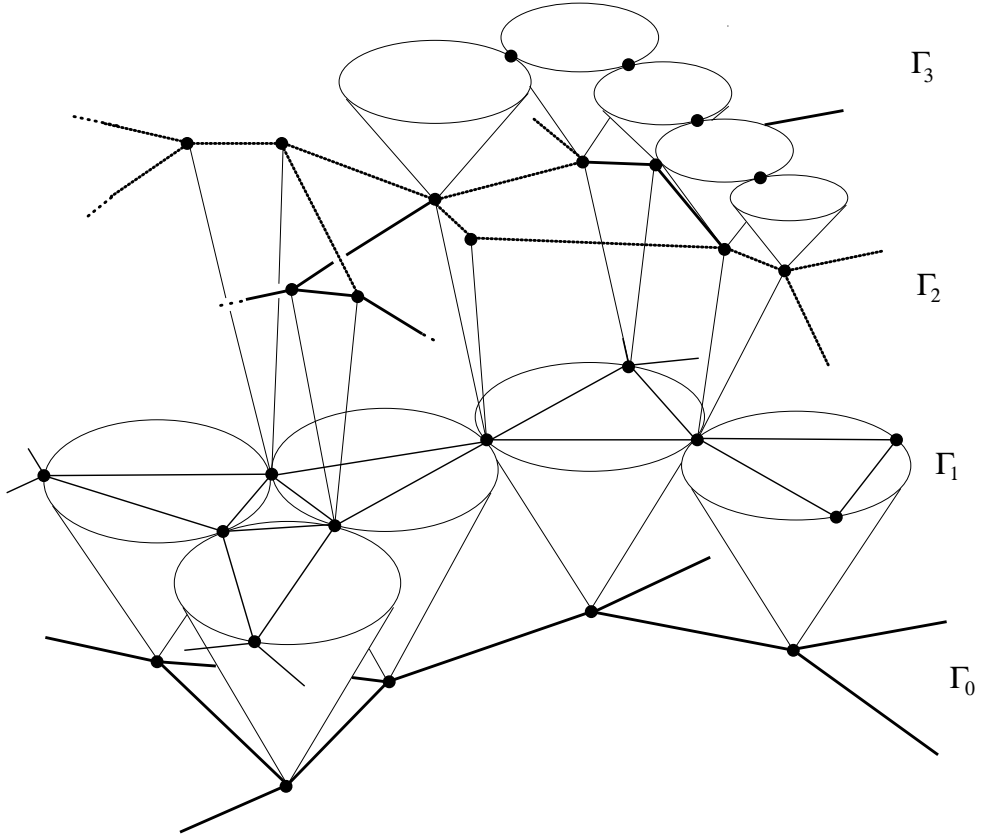}}}
\caption{A piece of a spacetime network with the fourth level under
construction.  The lightcones symbolize the null edges 
formed by Rule 1.}
\label{stnet}
\end{figure}

Thus we have achieved our goal, which is an amplitude for
evolution of spinnetworks in terms of a sum over intermediate
four dimensional ``spacetime nets", whose construction defines
a discrete version of causal structure that is then obeyed by the
rule for assigning amplitudes.

\subsection{Interpretation in terms of timelike surfaces}

The  
spacetime nets $\cal G$ defined by the evolution rules contain
sets of ``timelike" surfaces.  Examples of these are triangles defined
by Rules 1 and 2, which we see shaded, respectively, 
in Figures 4 and 10, and a set of squares which
are created by the evolution of edges that are preserved
by Rule 2.    The former have two null edges and
one spacelike edge while the latter have two of each.  
The labelings may be extended to labelings of these timelike surfaces
by assigning to each triangle or square the labelings of its
spacelike edges (the two spacelike edges of each square have the
same labels.)    This creates two kinds of surfaces that share
common labels: diamonds with four null edges and 
hexagons with six null edges.
These timelike surfaces may 
be considered to be the primary objects out of which the
spacetime nets are constructed.   The result is a theory in
which the amplitude for a spin network to evolve to another
one is given by a sum over terms, each of which consists of
a set of labeled timelike surfaces.  This is the same as in
the formulation of Reisenberger and Rovelli \cite{RR} and
in topological quantum field theory \cite{4dtqft}.

\section{The $3+1$ dimensional theory}

We can raise the spatial dimension from $2$ to $3$ by making
several modifications in the structure we just defined.

\begin{itemize}

\item{} We raise the valence of each node of the initial spin
network $\Gamma_0$ from $3$ to $4$.  Most
combinatorial four valent graphs are not planar, but they can
each be embedded in a three dimensional manifold.  

\item{}Each node of the initial four valent network must now
be labeled,  by inserting a virtual trivalent graph
as described in \cite{sn1}.    
Furthermore, if we consider possible embeddings
of $\Gamma$ into a three manifold $\Sigma$, 
generic nodes will contribute to the
volume of $\Sigma$.  

\item{}The successor network constructed by Rule 1 will
consist of six valent nodes (Figure \ref{sixvalent}).  The complete
graphs, corresponding to the 
triangles of the $2+1$ theories, are now tetrahedra.  There
is thus one tetrahedra ${\cal T}_n$ in $\Gamma_1$
corresponding to each node $n$ of $\Gamma_0$.  
Associated
to each such node of $\Gamma_0$ there is  then a
four simplex, consisting of the spacelike tetrahedra
${\cal T}_n$ in $\Gamma_1$ just described and the 
four null lines that connect its nodes to $n$.  We shall
call this ${\cal S}_n$.

As before, the null lines in ${\cal S}_n$ are labeled by the
spins of the $4$ edges incident on $n$.  $n$ is also labeled.
We need to prescribe an amplitude for each assignments
of labelings to the four nodes and six  edges of
${\cal T}_n$.  This will be called
${\cal J}^{15}$ as it is a function of $15$ labels, corresponding
to ten edges and five nodes of ${\cal S}_n$.

The dynamics are then given by a choice of ${\cal J}^{15}$.
As in the $2+1$ dimensional case, the space of such
functions is the parameter space of the theory.
The simplest choice is to take ${\cal J}^{15}$ 
equal to the $15j$ symbol\cite{4dtqft}, which, as in the
$2+1$ dimensional case, is associated with topological quantum
field theories.

\item{}To complete the specification of Rule 1 we must
say how the new six-valent nodes created are labeled.
There is a natural prescription associated with this that
preserves the principle of causality.
Each new six valent node $n^\prime_{12}$ may be virtually split into
two four-valent nodes, joined by a virtual edge that is
parallel to the edge $e_{12}$ it came from.  Let the label
of $e_{12}$ be $m$.
This edge joins two nodes, $n_1$ and
$n_2$.  Each is in a state $|\Psi_1\rangle $ and 
$|\Psi_2\rangle$ in their corresponding spaces
${\cal V}_4$, each of which may be describe by a superposition
of virtual trivalent graphs.  We may associate
to the node the state in the associated space ${\cal V}_{6}$
which is described by the states $|\Psi_1\rangle$ and $|\Psi_2\rangle$
associated with  $n_1$ and $n_2$ joined by a virtual
edge labeled by $m$.  That is, we simply read the
subgraph consisting of $n_1$, $n_2$ and $e_{12}$ as
describing a state in the space ${\cal V}_6$ associated
to the new node.

\begin{figure}
\centerline{\mbox{\epsfig{file=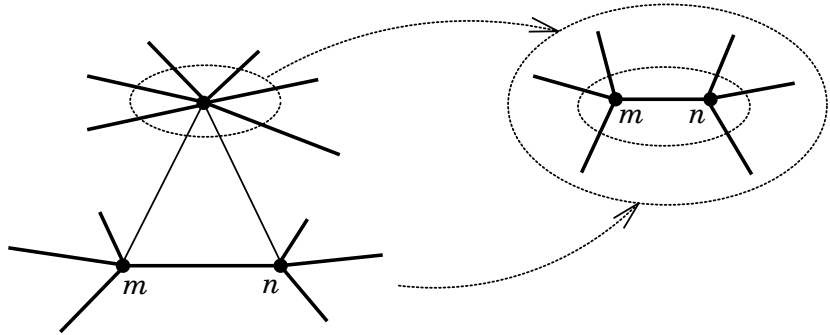}}}
\caption{The labeling of a six valent node in the $3+1$ theory:
The subgraph indicated is the state of the new node.}
\label{sixvalent}
\end{figure}

\item{}Rule 2 then must break each $6$ valent node
of $\Gamma_1$ back down into a pair of four valent
nodes.  As in the $2+1$ dimensional case, we will sum
over the different ways of doing this, with an amplitude
given by the inner product between the state $|m\rangle$ given
by the labeling on the six valent node, $m$,  and the
state given by the pair of four valent nodes in $\Gamma_2$,
with their labelings.  There are $20$ ways to make the split,
each of which produces a pair of four valent nodes, each
separated by new edge.  Each of the two four valent edges
must then be labeled as well; a basis of states here must
be labeled by a virtual edge.   Associated with each way
of splitting the six valent node we then have a state
in its ${\cal V}_6$, which we may call $|p\rangle$.  The
(unormalized) amplitude for each split will then be
given by $\langle p|m\rangle$.

\end{itemize}

By comparing the two sets of rules we can see several reasons
why the first is associated with $2+1$ theory and the second with 
$3+1$.  1) In the first case the spatial spin networks are planar,
in the second, generally not.  2)  The spacetime objects which
represent the elementary discrete future null cones are 
three and four simplices, repsectively.  3)  The corresponding
simplest choices for amplitudes in each case are the
$6j$ and $15j$ symbols, which are the amplitudes associated
with the simplest versions of $3$ and $4$ dimensional
topological quantum field theory.  4) All the nodes of the 
$3+1$ dimensional theory
have non-zero quanta of three dimensional volume generically.

\section{$1+1$ dimensional models}

We have defined a family of discrete quantum theories
of gravity in $2+1$ and $3+1$ dimensions, each of which is
described by a choice of a function of the labelings on a
tetrahedron or four simplex, respectively.
Given a choice of these functions, we have 
a complete perscription for
a path integral for quantum gravity.  However, it is not simple to
work out its consequences and investigate questions such as
the existence of a classical limit.  It is then useful to construct
an analogous model in 1+1 dimensions, which may be analyzed
more easily.   

\subsection{A first $1+1$ dimensional model}

\begin{figure}
\centerline{\mbox{\epsfig{file=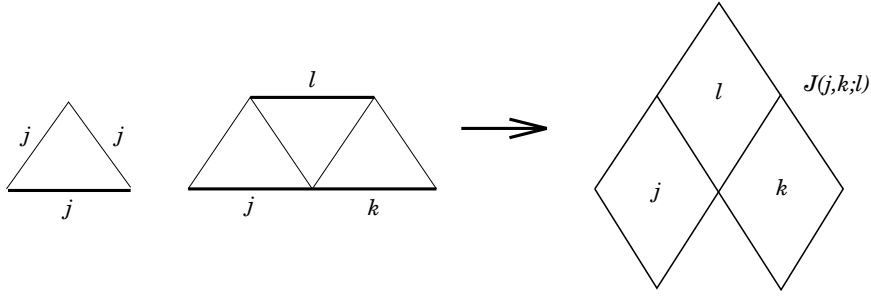}}}
\caption{A $1+1$ dimensional model.}
\label{1+1}
\end{figure}

A spatial state of this consists of a 
circle broken up into $N$ segments by $N-1$ nodes.  
The segments are labeled  by elements
of some set of colors $\cal L$.  These will not be interpreted as
spins, as then conservation of angular momentum would restrict
them to be all the same (See Figure \ref{1+1}). 

In this case there is only one evolution rule, which is essentially
Rule 1 of the higher dimensional case.  Each node emits
two null edges, one going to the left and one to the right.
Each edge $e_i$ then gives rise to a new node $n^\prime_i$.
This  can be interpreted as the event where the right moving 
light ray
from the left edge of $e_i$ meets the left moving light ray from
its right edge.  Each old node is then replaced by a new edge
$e_i^\prime$, with a new labeling $l$. (See Figure \ref{1+1}).  There
must be a rule which gives an amplitude $J(j,k;l)$ for the
new edge $e_i^\prime$, given the labelings of the two edges that
were adjacent to the node that gave rise to it.  

The resulting $1+1$ dimensional spacetime net can also be
interpreted in terms of labeled timelike surfaces, which in this
case are all diamonds.  The amplitude $J(j,k;l)$ is then
assigned to each triple of neighboring diamonds, in which
the $j$ and $k$ are the labels of the diamond just to the past
of the diamond labeled by $l$ (See Figure \ref{1+1}).

Given a choice of the amplitude $J(j,k;l)$ we then have a 
complete rule for the amplitude of the evolution of states.
Given an initial state $\Gamma_i$ and a final state $\Gamma_f$,
each given by a circle of labeled segments.  Then the amplitude
for the transition is given by
\f
{\cal A}_{i \rightarrow f} = \sum_{\cal G} \prod J(i,j;k)
\ff
where the sum is over all spacetime nets $\cal G$ whose
boundary $\partial {\cal G} = \Gamma_i + \Gamma_f$.
and for each spacetime net the product is over all triples
of nearest neighbor diamonds.

\begin{figure}
\centerline{\mbox{\epsfig{file=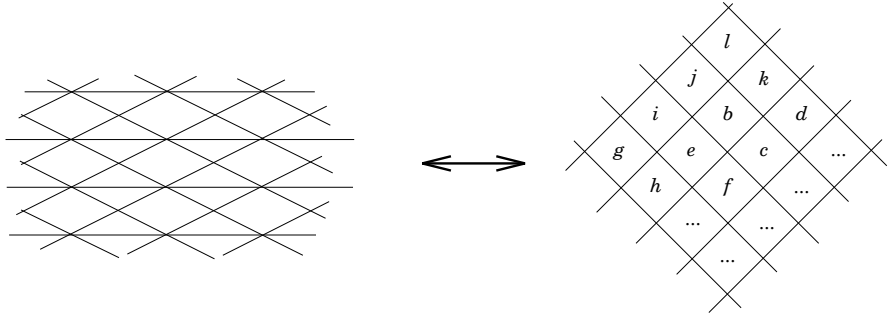}}}
\caption{  A history for the first kind of $1+1$ dimensional model.}
\label{1lat}
\end{figure}

We may give two examples of such nearest neighbor rules.
First, we can model the system as a lattice gauge theory.  In
this case $i,j,k,...$ are elements of some group $G$ and
$J(i,j;k)=exp [\imath \beta Tr_\rho [ijk ]]$ where the 
trace is taken in a representation
$\rho$.  The resulting theory is a kind of anisotropic $1+1$
dimensional lattice gauge theory.   

A second model is a kind of Potts model in which
$i,j,k \in Z_n$ and $J(j,k;l)=exp[\imath\beta |j+k-2l|^2] $.

\subsection{Directed percolation model}

\begin{figure}
\centerline{\mbox{\epsfig{file=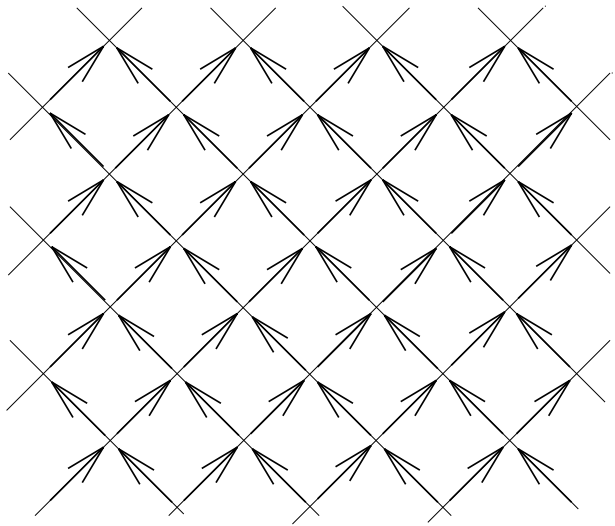}}}
\caption{The lattice of the percolation model.}
\label{perc}
\end{figure}

\begin{figure}
\centerline{\mbox{\epsfig{file=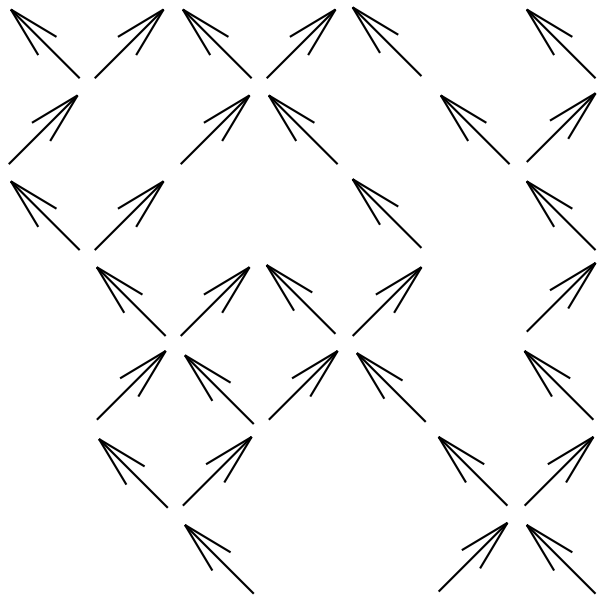}}}
\caption{A history in the percolation model.}
\label{2perc}
\end{figure}

We can describe a third kind of 
$1+1$ dimensional model by doing the following.
Represent each diamond as a site of a $1+1$ dimensional lattice,
and represent each future pointing causal link between diamonds
as a null edge between the corresponding nodes (see Figure \ref{perc}).
Call the resulting 
$1+1$ dimensional spacetime lattice, consisting of only of 
the nodes and the null edges, $\Delta$. (This is what we called
the internet before).  
There is a form of the
theory in which the 
dynamical variables  are  associated only with the null
edges.  In the simplest case, each null edge $l_{ij}$, connecting
nodes $i$ and $j$ is either on or off.  These states may be 
represented by either the presence of absence of an arrow.
  A history of the system is given by a choice
of on or off for each null edge in $\Delta$.  It may
be represented by a graph $\rho$  gotten by removing from
$\Delta$ those null edges which are off (see Figure \ref{2perc}).

The evolution rule will consist of an amplitude 
${\cal A}_n$ for each node $n$ which is a function of the
four null edges incident on $n$, the two that come from the
past and the two that go out to the future.    We will not put
any condition on these rules, except to impose that the amplitude
that one or both arrows in the future of a node $n$ are on must
be zero in the case that both incoming arrows are off.  A specification
of the amplitude ${\cal A}_n$ is a function of the sixteen possible
states of the two incoming and two outgoing arrows.

It is interesting to note that this theory is closely related to three
kinds of theories that have been studied before.
First, each history $\rho$ is a causal set.  We may then seek
to embed it in a $1+1$ dimensional manifold $\Sigma$ with metric
$g_{ab}$ preserving both the causal structure and the spacetime
volume (where we follow 'tHooft's suggestion that the 
spacetime volume
of a region is the number of nodes in Planck 
units\cite{thooft-cs}).

Second, this theory is closely related to a class of statistical
models known as directed percolation.   There are a number of
such models\cite{DP}.  In the simplest, each arrow may be on or off
with an independent probability $p$.  In more sophisticated
models, there are probabilities associated to each choice of
in arrows and out arrows, subject only to the restriction that
the probability for out arrows to be on is zero if all both in arrows
are off.  This model then corresponds to a special case of the
quantum gravity model we have described in which the 
amplitudes are all real and positive so that they sum
according to the rules of classical probabilities.  

Interestingly enough, both of these models have critical behavior,
which corresponds to percolation\cite{DP}.  While they are different,
they are in the same universality class, which is to say that their
critical behavior is identical.  A large number of other statistical
mechanics models
fall into the same universality class\cite{DP}.

The third kind of closely related theories are
binary networks, or cellular automata,
in which there is a definite rule by which
the outgoing arrows are determined as a function of the
incoming arrows.  This case is also a special case of
our model in which the amplitudes are treated as classical
probabilities.  Within this class of models are also some
that have critical behavior. There is even a class of such
models with {\it self-organized} critical behavior\cite{MZ}.

Directed percolation models exist and have non-trivial 
critical behavior for
all dimensions up to $3+1$.  It is interesting to note that
corresponding to each such theory one gets a statistical theory of
the corresponding causal sets.  Thus, this is a connection that
may be fruitfully pursued.  The most important question to
understand is whether the theories with complex amplitudes
rather than classical probabilities have critical behavior.  It
would be especially interesting if such theories were to have
self-organized critical behavior.

\section{Conclusions}

We have described here a class of theories for $1+1$, $2+1$ and
$3+1$ dimensional quantum gravity.  Each of them gives a
discrete path integral for the amplitude for any spin network
state $|\Gamma_{initial}\rangle$ to evolve to a final state
$|\Gamma_{final}\rangle$.  In each case the dynamics are specified
by giving a complex function, $J(i,j,...)$ of the labelings of a $d+1$
dimensional simplex.  The possible such functions consistent
with the symmetries of the simplex that do not mix spacelike and
null edges thus comprise the parameter space of this class of theories.
In $2+1$ and $3+1$ there are natural choices for these functions,
which are the normalized $6j$ and $15j$ symbols respectively.  We may
conjecture that these choices lead to a topological quantum field
theories, given the closeness of the theory to them in that case, but
this has yet to be shown.  What needs then to be done is to explore
the behavior of the theory given different choices for the
$J$'s.  Those choices that have critical behavior
will be candidates for quantum theories of gravity.  If they
exist they will be theories that are finite
and discrete at the Planck scale, are based on the kinematical structures
discovered by canonically quantizing general relativity and have a
continuum limit in which classical spacetime is reconstructed by making use
of the causal relations of the spacetime networks.  
In these cases the continuum limit should be
described by the Einstein's equations, in the limit of large radius
of curvature.

We believe that the crucial problem to be studied in this class of theories is
the existence of critical behavior, as this is necessary for the
existence of a continuum limit.  We find especially interesting 
the fact that the spacetime networks have two structures,
associated with spin networks and causal sets.  
Each theory by itself has failed to have enough
structure to ensure the existence of a good continuum limit, we hope
that by using both of them, it will be possible to investigate the
question of the existence of the continuum limit.

Another related issue is the possibility that the critical behavior of
these theories will be {\it self-organized}. At least philosophically,
this would be attractive as it would save us from the problem of having
to believe that the existence of a classical limit for a quantum
theory of gravity depends on the fine tuning of some parameters.

To investigate these kinds of questions, we find the relationship
with directed percolation models very promising.  In particular,
the fact that every configuration of a directed percolation model
is a causal set means that each directed percolation model is at the
same time a statistical theory of discrete spacetime geometry.  As
there exist such models in $d+1$ dimensions, for $d$ at least up to
$3$, this gives us a rich new class of models of dynamical spacetime
geometry, which are already set up to study the key problem of the
existence of the continuum limit. 

This connection raises further interesting questions, which deserve
study.  Among them are whether there is a universality class of
quantum directed percolation, in which the histories are weighed
by complex amplitudes rather than real probabilities.

A related line of attack is to construct a renormalization group
transformation on the space of such theories.  One approach, which
preserves both the kinematical interpretation of the spin network
states and the causal properties of the spacetime networks is
presently under development\cite{fl-rg}.

Other avenues of attack concern the relationship of the class
of theories we propose here with topological quantum field
theories in $3$ and $4$ spacetime dimensions.  As 
mentioned above, we suspect that
particular choices of the dynamical parameters, in which the
$J$'s are taken proportional to $6j$ or $15j$ symbols, respectively,
are closely related to the respective topological quantum field
theories.  If this is the case then it may provide an avenue of
attack on the renormalization group behavior of theories which
are close to the $TQFT$'s.  These theories are also likely
to be closely related to the Euclidean Reisenberger-Rovelli
models.

Finally, at the classical level the theories we describe may be
related to null strut regge calculus\cite{nullstrut}.

Before closing this paper we would like to elaborate on several
of these issues in more detail.

\subsection{Causal sets and the classical limit}

To see how 
the causal set structure may play a role in the classical limit
let us ignore for a moment the additional structure associated
with the spin networks.  Considering only its causal 
structure, a spacetime 
network consists of a set of
points together with causal relations.  These causal relations
are coded entirely in the null edges; the causal structure is
completely independent of the spinnets that tie together
nodes on a single ``spatial slice".  It is coded entirely in
what we called ``the internet".

Given any causal set, we can ask if it embeds in a spacetime
manifold, $\cal M$ with spacetime metric, $g_{ab}$, 
such that the nodes are mapped to points of $\cal M$ such
that the causal structure is preserved.  Even if it does the
metric and embedding will not be unique, but we can ask if
all metrics and embeddings of a given spacetime net
share an averaged casual structure.  That is, is there some
coarse grained metric $\tilde{g}_{ab}$ defined by averaging
the metrics $g_{ab}$ over many Planck volumes, such that
all embeddings of the spacetime net agree?  If so, then we
can say that the spacetime net has a classical limit, given
by the common average, $\tilde{g}_{ab}$.  

The spacetime metrics $g_{ab}$ will be defined up to 
conformal structure.   It follows that the averaged metric
will also be only determined up to conformal structure,
at least as long as the conformal transformations are
sufficiently slowly varying.  

To fix the conformal class of the averaged metric will 
require additional information.
This information has to do with the volume of spacetime
regions.  However, this information is likely provided by the
other information in the spin networks. First of all, the spatial
volumes of regions of spatial slices are fixed by the labels
of the nodes and edges of each spin network.  It remains to
fix the lapse functions.  However, these are not free, as
they are fixed by the causal relations. (The implications of
this are discussed below.)  Thus, it is likely that the spin networks
provide sufficient additional information to allow the reconstruction
of the conformal factor.  
The details of how the causal structure and spin network structures
may interact to determine an embedding in a classical geometry
remain to be worked out.  But it seems promising that by combining
the two structures we have  a possibility of a kind of classical
limit that is not available for Euclidean quantum gravity.  

\subsection{Time reversal invariance and noninvariance}

It is evident that the rules we have defined here are not time
reversal invariant, for example the number of nodes in each
spin network $\Gamma_i$ generally increases with $i$.  If it were
shown to be necessary, it would be possible to remedy 
this by simply including 
additional rules which are the time reversals of Rules 1 and 2.
However, there are several reasons to investigate the theory as
is.  As we have argued, if there is critical behavior the classical
Einstein's equations must govern the classical limit, so that time
reversal invariance may be restored at the classical level.  Time
reversal invariance may in fact play a role in the establishment of
the continuum limit, as it is a feature of directed percolation models
that seem related to those we study here.  For one thing, we would
like a satisfactory theory to have a continuum limit that was a
consequence of self-organized critical behavior, and such behavior
is normally found in the domain of non-equilibrium, time reversal
non-invariant systems.  Finally, there are a number of independent
arguments that suggest that quantum gravity either may be, or must
be time reversal non-invariant.

\subsection{Spacetime diffeomorphism invariance}

The reader may object that if the dynamics is not generated by
a constraint we have not imposed all of the gauge invariance of the
theory.  However, we may note that 
the rules we described do not allow a freedom to continuously
deform the spatial slices.  If we imagine that the spacetimenet
is embedded in a background $d+1$ dimensional spacetime
${\cal M},g_{ab}$ preverving the discrete light cone structure then
it will not be possible to vary the embeddings of the spatial
slices independently.  This is because, once the embedding of the
first slice is picked, the second, which is constructed by Rule 1,
is determined by the intersection of null cones.   

Thus, the rules we have given fix the gauge freedom corresponding
to the Hamiltonian constraint.  An important question is then
whether the full spacetime diffeomorphism invariance is recovered
as a gauge freedom of the effective action in the continuum limit.
As the relationship with the classical geometry will go through the
poset construction, it is likely that this is the case, as the 
correspondence between the exact discrete spacetimenet and the
approximate continuum description uses only the causal structure,
which is spacetime diffeomorphism invariant.

If this is the case, then the effective classical theory that
emerges will satisfy a Hamiltonian constraint, as that will
be a reflection of the invariance of the effective action under
spacetime diffeomorphisms.  Thus, the Hamiltonian constraint will
be recovered in the classical limit, even if it is not put in at the
fundamental level of the theory.

We may note that this implies that the discrete theory may have
a notion of time and evolution that is not coded in its continuum limit.
We believe that this possibility that discreteness may resolve
the problem of time in quantum gravity is significant, and should
be investigated in its own right.

\section*{ACKNOWLEDGEMENTS}

We are grateful to Per Bak, 
Roumen Borissov, Louis Crane, Chris Isham, Stuart Kauffman, 
Silvia Onofrei, Maya
Paczuski and Adam Ritz for comments and suggestions.  This work was 
supported by NSF grant PHY-9514240 to The Pennsylvania State
University, a NASA grant to The Santa Fe Institute, and by
the A. S. Onassis foundation.  FM would also like to thank
Abhay Ashtekar for hospitality at Penn State.

\end{document}